          [2001/04/25 1.1 (PWD)]
\documentclass[a4paper,twocolumn]{esapub}

\usepackage[square,numbers,sort&compress]{natbib}
\usepackage{graphicx}
\title{The fate of radiating black holes in noncommutative geometry}
\author[(1)]{Piero Nicolini}\author[(2)]{Anais Smailagic}
\author[(3)]{Euro Spallucci}

\affil[(1)]{{\it Dipartimento di Matematica e Informatica,
Universit\`a  di Trieste, via Valerio 12/b, I-34127 Trieste
(Italy)

Istituto Nazionale di Fisica Nucleare, Sezione di Trieste, I-34014
Trieste (Italy)

Dipartimento di Matematica, Politecnico di Torino, I-10129 Turin
(Italy)

Institut Jo\v{z}ef Stefan, 1000 Ljubljana (Slovenia)

E-mail: nicolini@cmfd.univ.trieste.it}}

\affil[(2)]{{\it Istituto Nazionale di Fisica Nucleare, Sezione di
Trieste, strada costiera 11, I-34014 Trieste (Italy)

E-mail: smailagic@ictp.trieste.it}}

\affil[(3)]{{\it Dipartimento di Fisica Teorica, Universit\`a di
Trieste, strada costiera 11, I-34014 Trieste (Italy)

Istituto Nazionale di Fisica Nucleare, Sezione di Trieste, I-34014
Trieste (Italy)

 E-mail:
spallucci@trieste.infn.it}}

\begin{document}


\maketitle

\begin{abstract}
We investigate the behavior of a  radiating Schwarzschild black
hole toy-model in a $2D$ noncommutative spacetime. It is shown
that coordinate noncommutativity leads to: i) the existence of a
minimal non-zero mass to which black hole can shrink; ii) a finite
maximum temperature that the black hole can reach before cooling
down to absolute zero; iii) the absence of any curvature
singularity. The proposed scenario offers a possible solution to
conventional difficulties when describing terminal phase of black
hole evaporation.
\end{abstract}

\section{Introduction}
 The theoretical discovery of radiating black holes
\citep{hawking} disclosed the first physically relevant window on
the mysteries of quantum gravity. After thirty years of intensive
research in this field  various aspects of the problem still
remain under debate (~see \citep[]{paddy} for a recent review with
an extensive reference list ~). For instance, a fully satisfactory
description of the late stage of black hole evaporation is still
missing. The string/black hole correspondence principle
\citep{corr} suggests that in this extreme regime stringy effects
cannot be neglected. This is just one of many examples of how the
development of string theory has affected various aspects of
theoretical physics. Among different outcomes of string theory, we
focus on  the result that target spacetime coordinates become
\textit{noncommuting} operators on a $D$-brane \citep{sw,sw2}.
Thus, string-brane coupling has put in evidence the necessity of
\textit{spacetime quantization}. This indication gave a new boost
to reconsider older ideas of similar kind pioneered
in a, largely ignored, paper by Snyder \citep{sny}.\\
The noncommutativity of spacetime can be encoded in the commutator

\begin{equation}
\left[\, \mathbf{x}^\mu\ , \mathbf{x}^\nu\, \right]= i \,
\theta^{\mu\nu} \label{ncx}
\end{equation}

where $\theta^{\mu\nu}$ is an anti-symmetric matrix which
determines the fundamental cell discretization of spacetime much
in the same way as the Planck constant $\hbar$ discretizes the
phase space.\\
The modifications of quantum field theory implied by (\ref{ncx})
has been recently investigated to a large extent. Currently the
approach to noncommutative quantum field theory follows two
distinct paths: one is based on the Weyl-Wigner-Moyal
$\ast$-product and the other on coordinate coherent state
formalism \citep{ae0,ae2}. In a recent paper, following the second
formulation, it has been shown that controversial questions of
Lorentz invariance and unitary \citep{casino,casino2}, raised in
the $\ast$-product approach, can be solved assuming
$\theta^{\mu\nu}= \theta\, \mathrm{diag}\left(\, \epsilon_{ij}\ ,
\epsilon_{ij} \dots \,\right)$\citep{ae}, where $\theta$ is a
constant with dimension of length squared. Furthermore, the
coordinate coherent state approach profoundly modifies the
structure of the Feynman propagator rendering the theory
ultraviolet finite, i.e. curing the short distance behavior of
pointlike structures \cite{Nicolini 2004}. It is thus reasonable
to believe that noncommutativity could as well cure divergences
that appear, under various forms, in General Relativity.  In such
a framework, it would be  of particular interest to study the
effects of noncommutativity on the terminal phase of black hole
evaporation. In the standard formulation, the temperature diverges
and an arbitrarily large curvature state is reached. One hopes
that noncommutativity can have important consequences both on the
black hole thermal radiation and on the curvature singularity at
its
center. \\
In order to tackle this difficult problem, we are going to
consider a $2D$ toy-model where the effects of noncommuting
coordinates are
implemented through a modified Schwarzschild metric \citep{ag}. \\

\section{$2D$ black hole}
The most direct way to study black hole radiation is by using
$t$-$r$ section of the Schwarzschild vacuum solution
\citep{swave,swave2}

\begin{equation}
ds^2 = -\left(\, 1 +2\phi(\,r\,)\, \right)dt^2 + \left(\, 1
+2\phi(\,r\,) \right)^{-1}\, dr^2 \ ,\label{schw}
\end{equation}

where, $\phi(\,r\,)$ is the Newtonian potential

\begin{equation}
\phi\left(\,r\,\right)= \frac{M\, G_N}{ r}
\end{equation}

which solves the classical Poisson equation for a point-like
source described by Dirac delta-function

\begin{eqnarray}
&&\vec\nabla^2\, \phi = 4\pi\,G_N \,\delta\left(\, r\,\right)
 \label{poiscl}
\end{eqnarray}



In order  to incorporate noncommutative effects in the analysis of
black hole evaporation, at first glance, one should think of
modifying the $4D$ Einstein action and try to solve corresponding
field equations.  So far, a proper modification of this kind is
not known and one cannot proceed in this direction. At this point
we would like to reflect on the connection between
noncommutativity and curvature. Curvature on its own is a
geometrical structure defined over an underlying manifold. It
measures the strength of the gravitational field, i.e. is the
response to the presence of a mass-energy distribution,  and in
this context, one can speak of strong/weak-field regime. On the
other hand, noncommutativity is an
 \textit{intrinsic property} of the manifold itself even in \textit{absence
 of gravity}.  Thus, noncommutative modification of Schwarzschild,
 once introduced at a given curvature, will remain valid in any other
 field strength regime. This is exactly the gravitational analogue of
 the NC modification of quantum field theory  where
 the strength of the field is not an issue \citep{ae0}, \citep{ae2}.  \\
 Therefore, we propose  to start from the Poisson equation (\ref{poiscl}) which
 has a precise physical meaning
 and guarantees physical ground for noncommutative effects. Noncommutative
position operators are subject to the uncertainty principle
following from the commutator (\ref{ncx}). The physical effect of
noncommutativity is that the very concept of point-like object is
no more meaningful  and one has to deal with \textit{smeared}
objects only. These physical effects are discussed  in the
framework of
 coherent state approach
in \citep{ae}.  The implementation of these physically sound
arguments
  within the mathematical formalism
amounts to substitution of position Dirac-delta, characterizing
point-like structures, with Gaussian function of minimal width
$\sqrt{\theta}$ describing the corresponding smeared structures.

 \begin{eqnarray}
&&\vec\nabla^2\, \phi =
4\pi\,G_N \,\rho_\theta\left(\,\vec{x}\,\right)\ ,\nonumber\\
&&\rho_\theta\left(\,\vec{x}\,\right)=
\frac{M}{\left(\,2\pi\theta\,\right)^{3/2}}\,
\exp\left(-\vec{x}^{\,2}/4\theta\,\right) \label{poisson}
\end{eqnarray}

where, $M$ is the mass of the source, $G_N$ is the Newton constant
and $\rho_\theta$ is the Gaussian mass density.\\
Our strategy to construct a NC version of (\ref{schw}) is
summarized in the following scheme: \vskip1truecm

\begin{tabular}[width=5cm]{|lccc|}
\hline
 & & &  \\
GR: & $ g_{00}=1+2\phi $ & $\longrightarrow$ & $ \vec\nabla^2\,
\phi =
4\pi\,G_N \,\delta\left(\, r\,\right) $  \\
& & &  \\
$\updownarrow$ (?) & $\uparrow$ $\theta\to 0$ & & $\theta\ne 0$
$\downarrow$ $ \uparrow $
$\theta\to 0$\\
& & & \\
NCGR: & $ g_{00}=1+2\phi $ & $\longleftarrow$ & $ \vec\nabla^2\,
\phi =
4\pi\,G_N \,\rho_\theta\left(\, r\,\right) $ \\
& & & \\
\hline
\end{tabular}
\vskip1truecm

This is the best we can do until a fully self-consistent
reformulation of General Relativity over a noncommutative manifold
becomes available. We obtain the NC version of (\ref{schw}) as

\begin{equation}
 ds^2 = -\left(\, 1- \frac{2M}{r\sqrt{\pi}}\, \gamma \right)dt^2 + \left(\, 1-
\frac{2M}{r\sqrt{\pi}}\, \gamma \right)^{-1}\, dr^2 \label{ncs}
\end{equation}

where $\gamma\equiv\gamma\left(1/2 \ , r^2/4\theta\, \right)$ is
the lower incomplete Gamma function, with the definition

\begin{equation}
\gamma\left(1/2\ , r^2/4\theta\, \right)\equiv
\int_0^{r^2/4\theta} dt\, t^{-1/2} e^{-t}
\end{equation}

The line element (\ref{ncs})  describes the geometry of a
noncommutative black hole toy-model in $2D$ \footnote{In formula
(\ref{ncs}) we introduced geometrical units $G_N=c=\hbar=\kappa_B=
1$.} and should give us useful
indications about  possible noncommutative effects Hawking radiation.\\
The classical Schwarzschild metric
is obtained from (\ref{ncs})  in the limit $r/2\sqrt{\theta}\to\infty $.\\
The event horizon radius $ r_H$ is defined by the vanishing of
$g_{00}$. In our case it leads to the implicit equation

\begin{equation}
r_H= \frac{2M}{\sqrt{\pi}}\,\gamma\left(1/2\ , r^2_H/4\theta\,
\right) \label{horizon}
\end{equation}

Equation (\ref{horizon}) can be conveniently rewritten in terms of
the upper incomplete Gamma function as

\begin{equation}
 r_H= 2M\,\left[\, 1 -\frac{1}{\sqrt\pi}\, \Gamma\left(1/2\ ,
r^2_H/4\theta\, \right) \,\right] \label{horizon2}
\end{equation}

The first term in (\ref{horizon2}) is the Schwarzschild radius,
while the second term brings in $\theta$-corrections. \\
In the ``large radius'' regime $r^2_H/4\theta>>1$ equation
(\ref{horizon2}) can be solved by iteration. At the first order
approximation, we find

\begin{equation}
r_H= 2M\,\left(\, 1 - \sqrt{\frac{\theta}{\pi}}\frac{1}{M}\,
e^{-M^2/\theta}\,\right)
\end{equation}

The effect of noncommutativity is exponentially small, which is
reasonable to expect since at large distances spacetime can be
considered as a smooth classical manifold.
 On the other hand, at short distance one expects significant changes due to
the spacetime fuzziness, i.e. the ``quantum geometry''  becomes
important $r_H\sqrt\theta << 1$.  To this purpose it is convenient
to invert (\ref{horizon2}) and consider the black hole mass $M$ as
function of $r_H$:

\begin{figure}[h]
\begin{center}
\includegraphics[width=5cm,angle=270]{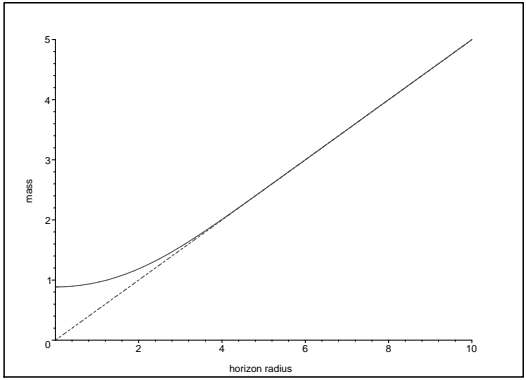}
\caption{\label{fig1}Mass vs horizon relation. In the commutative
case, dashed line, the mass is the linear function $M=r_H/2$
vanishing at the origin, while in the noncommutative case, solid
line, $M\left(\, r_H=0\,\right)=M_0$, i.e. for $M< M_0$ there is
no event horizon.}
\end{center}
\end{figure}

\begin{equation}
\frac{2M}{\sqrt\pi}= \frac{r_H}{\gamma\left(1/2\ , r^2_H/4\theta\,
\right)} \label{mh}
\end{equation}

In such a limit equation (\ref{mh}) leads to

\begin{equation}
M_0\equiv\lim_{r_H \to 0}\, M\left(\, r_H\,\right) =0.5\,
\sqrt{\pi\,\theta} \label{m0}
\end{equation}

which  is a new and interesting result. Noncommutativity implies a
minimal non-zero mass that allows the existence of an event horizon.\\
If the starting black hole   mass is such that $M> M_0$,  it can
radiate through the Hawking process until the value $M_0$ is
reached. At this point the horizon has totally evaporated leaving
behind a massive relic.
Black holes with mass $M< M_0$ do not exist. \\
The behavior of mass $M$ as a function of horizon radius is given
in Figure \ref{fig1}. To understand the physical nature of the
mass $M_0$ remnant, let us also consider the black hole
temperature as a function of $r_H$. It is given by

\begin{equation}
T_H\left(\, r_H \,\right)= \frac{1}{4\pi}\left[\,
\frac{1}{r_H}-\frac{\gamma^\prime\left(\, 1/2\ ;
r_H^2/4\theta\,\right)} {\gamma\left(\, 1/2\ ;
r_H^2/4\theta\,\right)}\, \right] \label{thnc}
\end{equation}

where the ``prime'' denotes differentiation with respect to $r$.\\

\begin{figure}[h]
\begin{center}
 \includegraphics[width=5cm, angle=270 ]{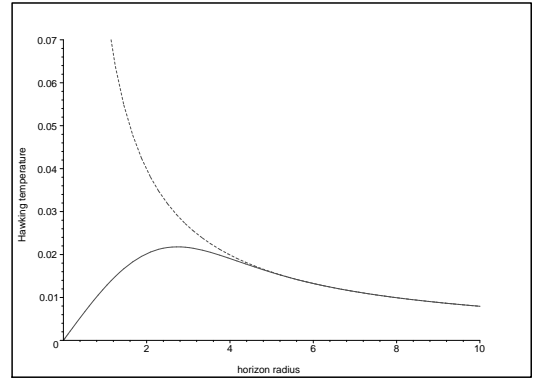}
\end{center}
\caption{\label{fig2} Hawking temperature $T_H$ as a function of
the horizon radius $r_H$. In the noncommutative case, solid curve,
one see the temperature reaches a maximum value $T_H^{Max.}=
2.18\times 10^{-2}/\sqrt{\theta}$ for $r_H= 2.74\sqrt\theta$, and
then decreases to zero as $r_H\to 0$. The commutative, divergent
behavior, dashed curve, is cured.}
\end{figure}

For large black holes, i.e. $r_H^2/4\theta>>1$, one recovers the
standard result for the Hawking temperature

\begin{equation}
T_H= \frac{1}{4\pi\, r_H} \label{th}
\end{equation}

At the initial state of radiation the black hole temperature
increases while the horizon radius is decreasing. It is crucial to
investigate what happens as $r_H\to 0$. In the standard
(~commutative~) case
  $T_H$ diverges and this puts limit on the
validity of the conventional description of Hawking radiation.
Against this scenario, formula (\ref{thnc}) leads to

\begin{equation}
T_H\sim \frac{r_H}{24\pi\theta} \ ,\qquad \hbox{as}\qquad r_H \to
0\label{zero}
\end{equation}

This is another intriguing result that has two important
consequences. Firstly, the emerging picture is that the black hole
has reached zero temperature and  the horizon has completely
evaporated. Nevertheless, we are left with a frozen, massive, remnant. \\
Secondly, passing from the regime of large radius to the regime of
small radius, (\ref{th}) and (\ref{zero}), implies the existence
of a \textit{maximum temperature} which is confirmed by the plot
in Figure \ref{fig2}.
 The plot gives the value  $T_H^{Max.}=2.18\times 10^{-2}/\sqrt\theta $.
The temperature behavior shows that  noncommutativity plays the
same role in General Relativity as in Quantum Field Theory, i.e.
removes short distance divergences. The resulting picture of black
hole behavior goes as follows. For $M >> M_0$ the temperature is
given by (\ref{th}) up to exponentially small corrections, and it
increases, as the mass is radiated away. $T_H$ reaches a maximum
value at $r_H=2.74\, \sqrt\theta$, and
then decreases down to zero as $r_H$ goes to zero. \\
At this point, important issue of Hawking radiation back-reaction
should be discussed. In commutative case one expects relevant
back-reaction effects during the terminal stage of evaporation
because of huge increase of temperature \citep{backr89a,backr89b}.
In our case, the role of noncommutativity is to cool down the
black hole in the final stage. As a consequence, there is a
suppression of quantum back-reaction since the black hole emits
less and less energy. Eventually, back-reaction may be important
during the maximum temperature phase. In order to estimate its
importance in this region, let us look at the thermal energy $E=T$
and the total mass $M$  near $r_H=2.74\,\sqrt\theta $. From
(\ref{mh}) one finds $M\sim \sqrt\theta \, M_{Pl.}^2$. In order to
have significant back-reaction effect $ T_H^{Max}$ should be of
the same order of magnitude as $M$. This condition leads to the
estimate

\begin{equation}
\sqrt{\theta}\sim 10^{-1}\, l_{Pl.}\sim 10^{-34}\, cm
\label{stima}
\end{equation}

Expected values of $\sqrt{\theta}$ are above the Planck length
$l_{Pl.}$ and (\ref{stima}) indicates that  back-reaction effects
are suppressed even at $T_H^{Max}\approx 10^{18}\, GeV$.  For this
reason we can safely use unmodified form of the metric
(\ref{schw})
during all the evaporation process.\\
 As it appears,  at the final stage of
evaporation a mass $M_0$ is left behind. One would be tempted to
say that the black hole evaporation has produced a \textit{naked
singularity} of mass $M_0$.
 We are going to show  that  this is not the case. In a $2D$ effective
geometry all the curvature tensors can be written in terms of the
Ricci scalar $R$. In case of  metric (\ref{ncs}) the scalar
curvature turns out to be

\begin{figure}[h]
\begin{center}
\includegraphics[width=5cm,angle=270]{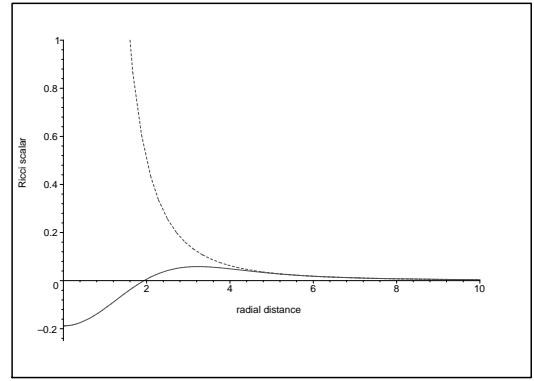}
\caption{\label{fig3} Ricci scalar as function of $r$. In the
noncommutative case, solid line, the curvature singularity in the
origin is removed, and $R\left(\,0\,\right)=-M/ 3\sqrt{\pi}
\theta^{3/2}$.}
\end{center}
\end{figure}

\begin{equation}
 R=\frac{2M}{\sqrt{\pi}}\left(\frac{2\gamma}{r^3}-
\frac{\gamma^\prime}{r^2}+
\frac{\gamma^{\prime\prime}}{r}\right)\label{ricci}
\end{equation}

One can check that at large distances (\ref{ricci}) reproduces the
usual $2D$ Schwarzschild scalar curvature

\begin{equation}
 R=\frac{4M}{r^3}
\end{equation}

In order to establish the geometrical picture of the frozen relic
we are going  to describe the behavior of the curvature
(\ref{ricci}) near $r=0$. In case of  naked singularity one should
obtain divergent curvature, $R\to\infty$. On the contrary, short
distance behavior of $R$ is

\begin{equation}
 R\simeq -\frac{M}{3\sqrt{\pi}\theta^{3/2}}\left(\, 1-
\frac{9}{20}\frac{r^2}{\theta}\, \right) \label{ricci0}
\end{equation}

For $r<<\sqrt\theta$ the curvature is actually \textit{constant
and negative}.

Regular black holes have been introduced as \textit{ad hoc} models
implementing the idea of a maximum curvature \cite{regbh}. On the
other hand we have found here an equivalent non singular $2D$
black hole as a solution of Einstein equations with a source
suitably prescribed by coordinate noncommutativity.\\

\section{Conclusions}
As a conclusion, the results derived in this work show that the
coordinate coherent state approach to noncommutative effects can
cure the singularity
problems at the terminal stage of black hole evaporation. \\
We have shown that noncommutativity is an intrinsic property of
the manifold itself and thus unaffected by the distribution of
matter. 
Matter curves a
noncommutative manifold in the same way as it curves a commutative
one, but cannot produce singular structures. Specifically, we have
shown that there is a minimal mass $M_0= 0.5\, \sqrt{\pi\theta}$
to which a black hole can decay through Hawking radiation. The
reason why it does not end-up into a naked singularity is due to
the finiteness of the  curvature at the origin. The everywhere
regular  geometry and the residual  mass $M_0$ are both
manifestations of the Gaussian de-localization of the source in
the noncommutative spacetime. On the thermodynamic side,  the same
kind of regularization  takes place eliminating the divergent
behavior of Hawking temperature. As a consequence there is a
maximum temperature that the black hole can reach before cooling
down to absolute zero. As already anticipated in the introduction,
noncommutativity regularizes divergent quantities in the final
stage of black hole evaporation in the same way it cured UV
infinities
in noncommutative quantum field theory.\\
We have also estimated that back-reaction does not modify the
original metric in a significant manner.

\section*{Acknowledgments}

One of us, P.N., thanks the ``Dipartimento di Fisica Teorica
dell'Universit\`a di Trieste'', the PRIN 2004 programme ``Modelli
della teoria cinetica matematica nello studio dei sistemi
complessi nelle scienze applicate'' and the CNR-NATO programme for
financial support.


\begin{thebibliography}{}
\bibliographystyle{aa}

\bibitem[Hawking(1975)]{hawking} Hawking S. W.,\ {\it Comm.\ Math.\
Phys.}
43, 199, 1975.

\bibitem[Padmanabhan(2005)]{paddy} Padmanabhan
T.,\ {\it Phys.\ Rep.} 406, 49, 2005.


\bibitem[Susskind(1993)]{corr} Susskind L.,\ {\it Phys.\ Rev.\ Lett.} 71, 2367,
1993.


\bibitem[Witten(1996),]{sw} Witten E., {\it Nucl.\ Phys.}\ B 460, 335,
1996.
\bibitem[Seiberg \& Witten(1999)]{sw2} Seiberg N. and  Witten E., {\it JHEP} 9909,
032, 1999.
\bibitem[Snyder(1947)]{sny}Snyder H. S., \ {\it Phys.\ Rev.}\ 71, 38,
1947.

\bibitem[Smailagic \& Spallucci(2003a)]{ae0} Smailagic A. and Spallucci E.,  {\it J.\ Phys.\ A} 36, L467,
2003.

\bibitem[Smailagic \& Spallucci(2003b)]{ae2} Smailagic A. and Spallucci E., {\it  J.\ Phys.\ A} 36, L517,
2003.

\bibitem[Chaichian et al.(2000)]{casino}Chaichian M., Demichev A. and Presnajder P.,
 {\it  Nucl.\ Phys.\ B} 567, 360, 2000.

\bibitem[Cho et al.(2000)]{casino2}Cho S., Hinterding R., Madore J. and Steinacker H.,\ {\it Int.\ J.\  Mod.\
Phys.\ D} 9, 161, 2000.

\bibitem[Smailagic \& Spallucci(2004)]{ae} Smailagic A. and Spallucci
E., {\it J.\ Phys. A} 37, 7169, 2004.

\bibitem{Nicolini 2004}  Nicolini P., \textit{Vacuum energy momentum
tensor in (2+1) NC scalar field theory,} arXiv:hep-th/0401204,
2004

\bibitem[Gruppuso(2005)]{ag} Gruppuso A., {\it J.\ Phys.\ A} 38,
2039, 2005.

\bibitem[Balbinot et al.(2001)]{swave2} Balbinot R., Fabbri A., Frolov V., Nicolini P. and Sutton
P.J. and Zelnikov A., {\it Phys.\ Rev.\ D }63, 084029, 2001.

\bibitem[Balbinot et al.(2002)]{swave} Balbinot R., Fabbri A., Nicolini P. and Sutton P.J.,
 {\it Phys.\ Rev.\ D}66, 024014, 2002.


\bibitem[Balbinot \& Barletta(1989a)]{backr89a} Balbinot R. and Barletta A., \ {\it Class.\ Quant.\
Grav.}
6, 195, 1989.

\bibitem[Balbinot \& Barletta(1989b)]{backr89b} Balbinot R.
and Barletta A.,  \ {\it Class.\ Quant.\ Grav. }6, 203, 1989.

\bibitem[Easson (2003)]{regbh}  Easson D. A., \textit{JHEP }0302:037, 2003; Hayward S.
A., \textit{Formation and evaporation of regular black holes,}
arXiv:gr-qc/0506126, 2005.

































\end{thebibliography}

\end{document}